\documentstyle[12pt,epsfig]{article}
\date{}
\begin{document}
\def\erf{\mathop{\rm erf}\nolimits}

\title{\normalsize{\bf{VOLUME CAPTURE AND VOLUME  REFLECTION OF ULTRARELATIVISTIC  PARTICLES IN
BENT SINGLE  CRYSTALS
}}}
\author{ Yu.A. Chesnokov, V.A. Maisheev and I.A. Yazynin\\
 \it{ Institute for High Energy Physics, 142281, Protvino, Russia }}
\maketitle
\begin{abstract}
Influence of volume capture on a process of volume reflection of ultrarelativistic
particles moving in bent single crystals was considered analytically. Relations
describing various distributions of particles involving in the process,
the probability of volume capture and behavior of channeling and dechanneling
fractions of a beam were obtained. Results of study will be useful
at creation of multicrystal devices for collimation and extraction of beams
on modern and future accelerators. 
\end{abstract}
\section{Introduction}
Volume reflection of charged particles in single crystals represents the coherent 
scattering of these particles by planar or axial electric fields of bent
crystallographic structures. For the first time, this effect was predicted in Ref.{\cite{TV}}
on the basis of Monte Carlo calculations. Recently in Ref. {\cite{MV}} an analytical description
of volume reflection of ultrarelativistic particles was considered.
Besides, this process was observed and investigated in a number of experiments \cite{Iv,Iva,WS}.
Results of study of volume reflection is interest to series of important applications
for accelerator technics (as beam collimation, extraction and others).
For work of these applications is assumed to use many times repeated volume reflection
process ( in  special crystal structures). One can expect that due to a high efficiency
of the process the sum efficiency of crystal devices will be also high enough.

The main process which influents on an efficiency of volume reflection is volume
capture \cite{CK} when moving in bent single crystals particles may be (due to multiple scattering
on atoms)  captured in a channeling regime. The analytical description  of volume reflection
(see \cite{MV}) gives good predictions for many characteristic of the process, but volume capture
was not considered in this theory.  

This paper is devoted to complete analytical consideration of volume reflection,
including influence of volume capture on the process. Main results of  Ref.{\cite{MV}} remain valid but 
their simple modernization is required. No doubt that Monte Carlo simulations give detailed information
about process at some specific initial conditions. However, an analytical description allows one to make  
observation of the problem as a whole. 

Our experience \cite{WS1} shows that an accuracy of the analytical calculations of such parameters
as a mean  and  mean square angles at volume reflection compares with the results of Monte Carlo simulations.  
Besides, in Ref.{ \cite{MV}} is shown how to use generalized parameters for description of the process.
Because of this, one can obtain set of results from one calculated result.
It is important for emulation of the process, when the experimental results are known
at some conditions (for example, for one energy of beam) and these results should be used at
another conditions (another energy). 

The paper is organized as follows. First, we describe our method for study of diffusion processes in bent single crystals.
Than based on this method we find the probability of volume capture and energy distribution underbarrier particles.
After that we investigate  behavior of the channeling fraction at its propagation in single crystal.
As result we get the analytical relations for  distribution functions of particles.

In the second part we illustrate the obtained relations by numerical calculations
and then we discuss and summarize presented in the paper results.

\section{One-trajectory approximation of a diffusion process}
Let us consider motion of the charged ultrarelativistic particle 
 in bent single crystal.
In the absence of  multiple scattering  we can write the following equation \cite{MV}:
\begin{equation}
E=E_0 \beta^2 {\dot x^2} /(2c^2) +U(x)+ \beta^2E_0 x/R,
\end{equation}
where $E_0$ is the particle energy, $E$ is the transversal energy, U(x) is the periodic interplanar
potential as a function of the transversal coordinate $x$, $R$ is the bending radius of a single crystal,
$\beta$ is the particle velocity divided on the velocity of light $c$.

Obviously, this equation reflects the conservation of the transversal energy.
Introduction  of  multiple scattering violates this condition. 
Let us suppose that the moving in bent crystal one separate particle is scattered on
the angle $\Delta \theta$ (by multiple scattering process). Then one can get
\begin{equation}
\Delta E =E_0\beta^2 \theta_m \Delta \theta+ {1\over 2} E_0(\Delta \theta)^2.
\end{equation}
Here $\theta_m \approx {\dot x} /c $ is the angle of particle due to regular motion in accordance with Eq.(1) 
Averaging over different values of $\Delta \theta$ we obtain the following equations
\begin{equation}
\Delta E={1\over 2} E_0(\Delta \theta)^2,
\end{equation}
\begin{equation}
(E^{'}-E- \bar{E})^2 =E_0^2\beta^2\theta^2_m (\Delta \theta)^2=  4(E-U(x)-\beta^2E_0 x /R) \Delta E
\end{equation}
where $E^{'}, \bar{E}$ are the variable value  and 
mean energy losses of the transversal energy.

After averaging over different angles for different particles we get distribution function
over changing transversal energy (from distribution function over plane scattered angle):
\begin{equation}
F(E^{'},l) = {1\over \sqrt{2\pi} \sigma_E } \exp{-[(E^{'} - E-\bar{E})^2 /(2\sigma_E^2)]},
\end{equation}  
where $\bar{E}$ and $\sigma_E$ are expressed through the mean square angle of
multiple scattering $\bar{\theta^2}$:
\begin{eqnarray}
\bar{E}={1\over 2}E_0 \bar{\theta^2}, \\
\sigma_E=E_0^2\beta^2\theta_m^2 \bar{\theta^2}.
\end{eqnarray} 

Note that function F satisfies to the following equation:
\begin{equation}
{\partial F \over \partial l}= 
-  {d \bar{ E}\over dl}  {\partial F\over \partial E^{'}}
+ {1\over 2}  { d\sigma^2_E \over dl} {\partial^2 F \over \partial {E^{'2}}}.
\end{equation}
Thus we got simple distribution function for one trajectory. This function depends
on only two parameters: $\bar{E}$ and $\sigma_E$. These parameters are independent of
the current transversal energy $E^{'}$. It is clear that obtained here approximation
is valid on a short distance, when  parameters $\bar{E}$ and $\sigma_E$ are
small enough with comparing of the characteristic value $\delta E$.

More traditional consideration of a similar problem is based on the diffusion equation with
the diffusion coefficients depending on the transversal energy.
This fact strongly complicate possibility for solution.

\section{ Volume capture probability}

Fig. 1 illustrates  behavior of the effective potential $U_e(x)=U(x)+\beta^2E_0 x/R$ as a function of the
transversal coordinate x.  The main problem of our consideration is determination of an inefficiency 
of volume reflection of charged particles in bent single crystals.  Mechanism of the inefficiency
is connected with the volume capture process in these structures. Really, the moving particle
has some probability to lose (because of multiple scattering) its transversal energy and to continue
motion in the channeling regime. However, the captured particle has some probability (due to
dechanneling) quickly return in underbarrier motion and to be similar as usual particle at volume reflection.
This fact requires the detailed description of the behavior of volume captured particles.  

In this section we find  the probability of volume capture. Fig. 1 illustrates this process.
Obviously that  volume capture takes place mainly in the vicinity of critical (reterned) point.
In Fig. 1 this region corresponds to $x_1 \le  x \le x_c$ for transversal energies of particle  $E_m \le E \le E_M$. 
In Ref.{\cite{MV}} was shown that  particles of  beam with the small angle divergence in  absence of multiple
scattering are equiprobably distributed over transversal energies.  Multiple scattering  changes the initial distribution
of particle over transversal energies. However, the particle distribution in every energy period $\delta E$  keeps approximately
equiprobable. Evidence is similar to considered case in Ref.{\cite{MV}}.

Multiple scattering we consider as in an amorphous medium but instead constant  nuclear density we use variable  one
in accordance with a current particle position. For this one can give different theoretical likelihood explanations
but more evident argument is fact that Monte Carlo calculations at the above mentioned assumptions give a good agreement with experiments.

Taking into account stated assumptions we define the probability of volume capture as an normalized number of particles (with the initial
energy range from $E_m$ till $E_M=E_m+\delta E$) which obtain (due to multiple scattering) the transversal energy
corresponding to finite planar motion (channeling).

As illustrated in Fig. 1 for every particle with the transversal energy
$E_m \le E \le E_M$ and coming to coordinate $x_1$ there are 
three different possibilities (see numbers near curves in Fig. 1):
1)to undergo reflection;
2)to lose the transversal energy and to become particle captured in
channeling regime;
3)to increase the transversal energy, so that $E> E_M$.
In the case of number 3   
theoretically the similar three possibilities is conserved but
for small enough bending radii the third variant is impossible practically.  
At condition of smallness of bending radius the probability for particle (with $E_m \le E \le E_M$)
to lose significantly the transversal energy before coordinate $x_1$ is very small.
Thus, we see that the volume capture probability one can represents
as a sum of the two terms $\varepsilon_1= \varepsilon_{1a}+ \varepsilon_{1b}$,
 where $\varepsilon_{1a}$ is the probability of volume capture in the
coordinate range from $x_1$ till $x_2$ and $\varepsilon_{1b}$ is the probability
in the coordinate range from $x_2$ till $x_3$.
       
Let us calculate the first term $\varepsilon_{1a}$. 
Our consideration will be based on one-trajectory approximation. Eq.(5) gives the distribution
for one trajectory of particle with the transversal energy $E$. The summary distribution function
is a result of integration over all the range of  transversal energies (from $E_m$ till $E_M$). 
Taking this into account we get the probability to change the transversal energy $E$ of a particle 
less than the critical energy $E_c$ in the following form:
\begin{equation}
\varepsilon_{1a} = {1\over 2} - {1\over 2} \int_0^1 
\erf({\xi + \xi_m \over \sqrt{2}\Sigma})d\xi,
\end{equation}
where $\xi_m(\xi) = \bar{ E} / \delta E$ and 
$\Sigma(\xi) =\sigma_E /\delta E$,
$\bar{E},\sigma_E$
are the mean and square mean losses of transversal energy (see Eqs.(6)-(7)), 
$\delta E =  E_0 \beta^2 d/ R$, 
$\erf(x) =2/\sqrt{\pi} \int_0^x \exp(-t^2)dt$. 
One can write the following equations
for  the mean and mean square  energy losses:
\begin{equation}
\bar{E} =2A^2(L_0/X_0)/E_0^2,
\end{equation}
\begin{equation}
\sigma_E =A\sqrt{2(L_1/X_0)}/E_0.
\end{equation}
Here $A \approx 10 \,-\,14 MeV$ is the constant value, $X_0$ is the radiation length
of a single crystal and $L_0$ and $L_1$ functions are:
\begin{equation}
L_0= {E_0c \over 2 \rho_0} \int_{t_1}^{t_2}
(\rho_a(x(t)) +\rho_e(x(t))/Z^2)dt,
\end{equation}
\begin{equation}
L_1= {2E_0c \over \rho_0} \int_{t_1}^{t_2}
(\rho_a(x(t)) +\rho_e(x(t))/Z^2) (E-U(x(t))-E_0 \beta^2x(t)/R) dt,
\end{equation}
where $\rho_a(x), \rho_e(x)$ are the planar atomic and
electron densities for a selected plane of single crystal, $Z$ is the atomic number,
 $\rho_0 \approx N_0/V_0$ is the mean atomic density ($N_0$ is number atoms in a
fundamental cell of volume $V_0$). $t_1$ is time which corresponds to
finding of particle in the nearest local maximum of U(x) before critical point $x_c$
(for this particle). Time $t_2$ corresponds to location of the particle in
a critical point. The coefficient 2 in Eqs. (11) and (12) is due to symmetry of particle motion
before and after a critical point.

Taking into account the relation:
\begin{equation}
 dt={dx \over \sqrt{{2c^2 \over E_0\beta^2} [E-U(x)-E_0 \beta^2 x/R]}},
\end{equation}
we can represent Eqs. (11) and (12) in the following form:
\begin{equation}
\bar{E} ={A^2 \over \sqrt{2E_0} \rho_0 X_0} 
\int_{x_1}^{x_c}
{(\rho_a(x) +\rho_e(x)/Z^2) \over \sqrt{E-U(x)-E_0 \beta^2 x/R}}dx,
\end{equation} 
\begin{equation}
\sigma_E={ 2^{3\over 4} A \over E_0^{1\over 4} (\rho_0 X_0)^{1\over 2} }
[\int_{x_1}^{x_c}
{(\rho_a(x) +\rho_e(x)/Z^2)  \sqrt{E-U(x)-E_0 \beta^2 x/R}}dx]^{1\over2}.
\end{equation}    
 It is convenient to  rewrite Eqs. (15) and (16) in the following more general form:
\begin{equation}
\bar{E}  ={A^2 d \over \sqrt{2E_0 U_0}  \rho_0 X_0}
\int_{y_1}^{y_c}
{(\rho_a(y) +\rho_e(y)/Z^2) \over  \sqrt{\nu/\kappa -U(y)/U_0 - y/\kappa }}dy
={A^2 d\over \sqrt{2E_0 U_0}  X_0}f_1(\nu,\kappa), 
\end{equation}
\begin{eqnarray}
\sigma_E={ 2^{3\over 4} A U_0^{1\over 4} d^{1\over 2} \over E_0^{1\over 4} (\rho_0 X_0)^{1\over 2} }
[\int_{y_1}^{y_c}
{(\rho_a(y) +\rho_e(y)/Z^2)\sqrt{\nu /\kappa -U(y)/U_0-y /\kappa}\,dy]^{1\over2}}  
={ 2^{3\over 4} A U_0^{1\over 4} d^{1\over 2} \over E_0^{1\over 4}  X_0^{1\over 2} } f_2(\nu,\kappa),
\end{eqnarray}  
where $y=x/d$ and $U_0$ is the potential  barrier of plane  for a straight single crystal and
the parameter $\kappa=U_0 R /( E_0\beta^2 d) $. This dimensionless parameter one can represent also as
$\kappa= R/R_0$, where $R_0= \beta^2 E_0 d /U_0$ is the characteristic radius of volume reflection \cite{MV}.  
Another dimensionless parameter $\nu = E/\delta E$. The parameter $\nu$ is coupled with the $\xi$-parameter (see Eq. (9))
by the relation: $\xi =\nu-\nu_m= (E-E_m)/\delta E$, where $E_m$ is the nearest  local maximum of a transversal energy before
the critical point $x_c$ $( \nu_m=E_m/\delta E)$.  Note the functions $ \bar{E} (\xi)$ and $\sigma_E(\xi)$
are periodic functions of $\xi$-variable with the period equal to 1. 

The probability $\varepsilon_1$ one can presented as sum of two first terms of Taylor series in the vicinity of $\kappa_0$ :
\begin{equation} 
\varepsilon_1(\kappa)=\varepsilon(\kappa_0)   +{d\varepsilon_1\over d\kappa}(\kappa_0) \kappa_c  
(  {\kappa \over  \kappa_c} - {\kappa_0 \over \kappa_c}),
\end{equation}
where $\kappa_c =  U_0 R_c/ (E_0\beta^2 d) $ where $R_c =E_0/{\cal{E}}_{max}$ is the critical radius of channeling
(${\cal{E}}_{max}$ is the maximal value of the interplanar electrical field). Thus, $\kappa_c= U_0/({\cal{E}}_{max}d)$
and $\kappa/ \kappa_c - \kappa_0/\kappa_c =R/R_c-R_0/R_c$ (radius $R_0 $ corresponds to $\kappa_0$-value).  
The presentation of a bending radius in units of critical radius is convenient for consideration of the process
at different particle energies.

From Eq.(9) one can get
\begin{equation}
{d \varepsilon \over d\kappa}=
- {1\over \sqrt{2\pi}}  \int_0^1 {\xi_m \over \Sigma} ({f_1^{'} \over f_1} -{f_2^{'} \over f_2}  -{ \xi \over \xi_ {m} }({1 \over  \kappa } +{f_2^{'}\over f_2})) 
 \exp{(-{(\xi+\xi_m)^2/(2\Sigma^2)) }} d\xi,
\end{equation}
where $f_1^{'}, f_2^{'}$ is the corresponding derivatives (of $f_1, f_2$ - functions) over $\kappa$-parameter. 

Taking into account Eqs.(17 ) and (18) one can represent Eq. (20 )  in the following form:
\begin{equation}
{d\varepsilon_1\over d\kappa}(\kappa_0) \kappa_c = 
{AU_0^{1\over4} \over 2^{7\over4} \sqrt{\pi} E_0^{1\over 4} {\cal{E}}_{max} d^{1\over 2} X_0^{1\over2}} J(\kappa_0), 
\end{equation}
where 
\begin{equation}
J(\kappa_0)=-\int_0^1 {f_1 \over f_2} ({f_1^{'} \over f_1} -{f_2^{'} \over f_2}  -{ \xi \over \xi_ {m} }({1 \over  \kappa_0 } +{f_2^{'}\over f_2})) 
 \exp{(-{(\xi+\xi_m)^2/(2\Sigma^2)) }} d\xi.
\end{equation}

The distribution of scattered particles over the relative transversal energy $\xi^{'}=(E^{'}-E_m)/\delta E$ follows from integration of Eq.(7):  
\begin{equation}
{\cal{F}}_a(\xi^{'})={dN(\xi^{'})\over d\xi^{'}}=
{1\over \sqrt{2\pi}} \int_0^1 {d\xi \over \Sigma} \exp{-{(\xi^{'}-\xi-\xi_m)^2 \over 2\Sigma^2}}.
\end{equation}
This function describes the all scattered particles at $-\infty < \xi^{'} <\infty$. 
The case $\xi^{'} <0 $ corresponds to particles with $E^{'}< E_m$ and the
case $\xi^{'}> 1$ corresponds to particles with $E^{'}> E_M$. 
It is significant to note that here and below (in similar equations)  the values $\xi_m$ and $\Sigma$ 
are  functions of the integration variable. 
Integration of Eq.(23)
at condition $\xi^{'} <0 $ gives Eq.(9) for total number particles, and at
condition $\xi^{'} > 1 $ gives corresponding total number of particles with $E>E_M$:
\begin{equation}
\varepsilon_0 ={1\over 2}-{1\over 2}\int_0^1 \erf({1-\xi-\xi_m \over \sqrt{2}\Sigma}) d\xi.
\end{equation}
The distribution of captured particles (due to the additional process, see the trajectory 3 in Fig.1)  is
\begin{equation}
{\cal{F}}_b(\xi^{''})= {1\over \sqrt{2\pi}} \int_1^{\infty} {\cal{F}}_a(\xi^{'}) {\exp[-(\xi^{''}- \xi^{'} -\xi_m)^2/(2\Sigma^2)] \over \Sigma}d\xi^{'}.
\end{equation}
Here $\xi^{''} < 1$, $\xi^{'}= (E^{'}-E_m)/\delta E$, $\xi^{''}= (E^{''}-E_m)/\delta E$, where 
$E^{'}$ and $E^{''}$ are the corresponding transversal energies (see Fig. 1).
 Thus, the total normalized on unit  number of captured particles in this approximation is
$\varepsilon_{1} =\varepsilon_{1a}+ \varepsilon_{1b}$, where $\varepsilon_{1b} = \int_{-\infty}^1 {\cal{F}}_b d\xi^{''}$.  
Note our consideration of volume capture is valid for small enough bending radii of single crystals.
The range of applicability of present description we will study below.

\section{Channeling of the captured particles}
For some time the captured particles move in a planar channel, but due to multiple
scattering (in the channeling regime) these particles obtain transversal energies
which exceed the potential barrier and hence the part of them return in underbarrier motion. 
In this section we consider this process on the basis of one-trajectory diffusion
approximation.

For calculations we need to know the losses of transversal energy of a channeled particle.
For this aim we can use Eqs. (17)-(18). However, we should put the transversal energy $E_m-U_R < E < E_m$
at location of particles in the range $x_1 < x < x_c$ ($U_R=E_m- E_{ch}$, see Fig. 1).  At this condition a particle
will be in the channeling regime. At first, we find the mean and mean square  losses
of a transversal energy per one period of motion $ l_c$ ($l_c$ is a function of a transversal energy). Then we can get approximately
that these losses depend on the factor $l/l_c$ (see Eqs.(10)-(13)), where $l$ is the length of a single crystal. We think at $l/l_c > 1$
this approximation is good enough.  

Now we find the distribution function of channeling fraction:
\begin{equation}
{\cal{F}}_c(\xi^{''}, l)= {1\over \sqrt{2\pi}} \int_{\approx  -U_R/\delta E}^0 [{\cal{F}}_a(\xi^{'})+{\cal{F}}_b(\xi^{'})]
{\exp[-(\xi^{''}-\xi^{'}-\xi_{mc})^2/(2\Sigma^2_c)]\over \Sigma_c} d\xi^{'} 
\end{equation}  
where $\xi_{mc} = (l/l_c) \bar{ E} /\delta E$ and $\Sigma_c= (l/l_c) \sigma_ E/\delta E$
and the variables $\xi^{'}$ and $\xi^{''}$ are the similar to analogous ones in Eq(25). 
Note due to  periodicity of the process we  wrote $\xi^{'}$-value as the argument of ${\cal{F}}_b$-function (instead of $\xi^{'}+1$).   
Obviously  this equation is valid for small enough $l$-values while the captured particles have transversal energies
more than $E_m-U_R$. 

Integration of ${\cal{F}}_c(\xi^{''})$-function over $\xi^{''}$ (from $\approx  -U_R/\delta E$ till 0) gives
the normalized number of channeling particles $\varepsilon_2(l)$. 
\section{Angle distribution of scattered particles}

Relations describing the angle distributions of scattered particles in the process of volume reflection were
obtained in Ref.{\cite{MV}}. However, these relations do not take into account the volume capture.
If we want take into account this factor we should divide  all particles on two sorts.
The first sort is the particles which were not captured and the second one is the captured particles.
The number of particles of the first sort is $1-\varepsilon_1$ and the number of particles of the second sort
is $\varepsilon_1$. The summary angle distribution of particles may be represented as 
$dN_T/ d\alpha(\alpha) =  dN_{vr} /d\alpha  (\alpha) + dN_{vc}/d\alpha(\alpha)$, where $dN_{vr} /d\alpha$
is the angle distribution of pure volume reflected particles (see detail description in Ref.{\cite{MV}}) and $dN_{vc} /d\alpha$ is the angle distribution of
volume captured particles. 

Let us consider the  bent single crystal of the finite thickness $L$. Then the volume captured particles we can 
represent as a sum of the two fraction: the first fraction $N_{ch}$ is the particles which enter from a single crystal 
in the channeling
regime (overbarrier motion) and the second one $N_{de}$ is the  particles which transmitted from channeling in
underbarrier motion (due to dechanneling mechanism). Let us suppose that normalized  number particles
of the first fraction is $\varepsilon_2$, then the normalized number of particles in the second fraction is equal to $\varepsilon_1-\varepsilon_2$.  

In Ref.{\cite{MV}} the term "physically narrow angle distribution of entering particles" was introduced. This term means
that characteristic size of angle distribution of the entering particles" exceeds the angle period $\delta \theta= d/(R\theta)$
(here $\theta$ is initial angle of particle) and particles are uniformly distributed within this period.
To this point we consider the multiple scattering only in narrow range of transversal coordinate $(< d$, see Eqs.(17) and (18)).
In this approximation we can write
\begin{equation}
{dN_{de}\over d\alpha}(\alpha) ={d\varepsilon_2\over d\varphi}(-\alpha_m/2+\varphi)\vartheta(\varphi_{max}-\varphi)\vartheta(\varphi ). 
\end{equation}
 here $\alpha_m$ is the mean angle of volume reflection, and angle $\varphi = (l-L/2)/R$, where 
  $\varphi_{max}=L/(2R)$, and $\vartheta(x) =1$ or $0$ at $x>0$ and $x<0$), respectively. 
This equation is valid for symmetric case of orientation of a single crystal.  Note the case of nonsymmetric
orientation is considered analogously. 

Fig. 1b  illustrates the geometry for Eq.(27). Here the broken line COD represents motion of a particle due to volume reflection.
The lines  CE and OD are the directions of particle motion before and after process. The angle DOE is
the mean angle of volume reflection. AB is a tangent line in the critical point.
We define the direction along  CE-line as zero angle  ( the initial direction of  particle motion).  
It means that the angle between OD- line (direction of volume reflection) and
the current direction of channeling flux is approximately equal to $ \alpha =-\alpha_m/2 +\varphi$ (see Eq.(27)). 

For final result we should take into account multiple scattering in the body of a single crystal.
In accordance with Ref.{\cite{MV}} we get
\begin{equation}
\langle {dN_{de}\over d\alpha}(\alpha)\rangle =\int_{-\infty}^{\infty} \rho(\varphi, \sigma_{ms}){dN_{de}\over d\alpha}(\alpha-\varphi)d\varphi 
\end{equation} 
where $\rho(\varphi,\sigma_{ms})$ is the distribution function of multiple scattering with $\sigma_{ms}$ corresponding
to thickness $L/2 -(l-L/2)\vartheta(l-L/2 ),\, l=\varphi R-L/2$  ($l-L/2$ is the part of  particle path in the body of single crystal
 which corresponds to motion 
in the channeling regime).

In our model the process of volume capture takes place in the transversal space
from $x1$ till $x3$ (see Fig. 1). It corresponds to $\delta$-function distribution of the captured particles. 
In reality, multiple scattering distributed this process 
in some transversal area and hence in some longitudinal space. One can describe the evolution of
distribution of volume captured particles with the help of the function \cite{MV}:
\begin{equation}
{dN_{ca}\over dl}={\varepsilon_1 \over \sqrt{2\pi} \sigma_{ca}R}\exp{-{(L/2-l)^2\over 2\sigma^2_{ca}R^2}}
\end{equation}  
where $\sigma_{ca}$ is the mean squared angle of multiple scattering
 corresponding to approximately half of single crystal thickness.

The density of dechanneling particles as a function of the length $l$ (or the angle $\varphi$ ) is described by the
relation as Eqs.(27) and (28) (at the condition $\alpha_m=0$) but for $\sigma_{ms}$ should be taken for a length equal to $L/2$. 
It means that we consider particles in the vicinity of local variable $l$ and
do not taken into account further motion in the body of a single crystal.
Let us denote this function as $dN_{de0}/dl$ (or $dN_{de0}/d\varphi = R dN_{de0}/dl$).

Now we can describe the behavior of channeling fraction as a function of the length (or the angle $\varphi$ )
by the following equation:
\begin{equation}
 \varepsilon_2(l)= {N_{ch}} (l) = \int_0^l ({dN_{ca}\over dl}(z)-{dN_{de0}\over dl}(z)) dz 
\end{equation}

Now we can find the angle distribution of channeling fraction.
It easy to get  the velocity ($v(t)= \dot x(t)$) distribution function (normalized on 1) of channeled particles for one 
fixed transversal energy $E$:
\begin{equation}
{dN \over dv}(v)  = {E_0 \over \tau c^2  |\beta^2 E_0/R -{\cal{E}}(x(t))|}
\end{equation}
where $\tau$ is the half of period of motion for channeled particle 
 and ${\cal{E}}$ is the intensity of planar electric field.
Taking account that $\theta \approx v/c$ we can recalculate this distribution
in angle one.
Knowing the distribution function over the transversal energy of channeled particles (see Eq.(26) ) we can find the result distribution
of these particles at the exit from a crystal. 

\section{Examples of calculations}
\subsection{Introduction remarks}
In this section in accordance with obtained equations we present calculations 
of volume reflection process, which taking account the processes of volume capture 
and channeling in bent single crystals. For illustration we selected some energies
of proton beams which approximately corresponds to energies of well-known
accelerators. The main and detailed calculations were done for silicon single
crystals, but some results were obtained and for some another single crystals.
For calculation of the potential, electric field, electron and atomic densities
in silicon we use values of atomic form factors from x-ray experiments.
The method of calculations one can find in Ref.{\cite{MV1}}. Note recent experimental
data \cite{WS1} show that the potential from x-ray diffraction gives more precise description
of measured characteristics of volume reflection for a silicon single crystal than Molier potential.    

\subsection{Energy losses over and under barrier particles}
For determination of different distribution functions we should know the
mean and mean squared energy losses of particles (see Eqs.(17)-(18)).
Fig. 2 illustrates these quantities for over and underbarrier motion
of 400 GeV/c protons. 
For these calculations we use $A= 11 MeV$ (see Eqs.(10)-(11)). This choice was
based on equation for the mean square angle of multiple scattering \cite{PP} $ \theta_{ms}=
13.6[MeV]/\beta cp \sqrt{l/X_0}[1+0.038\ln{(l/X_0)}]$, where $p$ is the particle momentum. 
For a small thickness $l$
$l/X_0 \approx 10^{-3}$ and more natural law $\theta_{ms} =A /E_0\sqrt{l/X_0}$
the best agreement is at $A=11 MeV$.

It is easy to recalculate the results of Fig. 2 for any energy of particles
(see Eqs.(17)-(18)). The universal functions $f_1$ and $f_2$  one can also find from Fig. 2.
  
\subsection{Simulations of volume capture}
The volume capture probability represents the sum of two terms $\varepsilon_1=\varepsilon_{1a}+\varepsilon_{1b}$.
Initially we consider behavior of the first term $\varepsilon_{1a}$.
Fig. 3 illustrates the calculations of this value for different proton energies.
One can see that at small values $R/R_c$ $\varepsilon_1$ is approximated good enough 
by linear function. From Eq.(21) we see that value $\kappa_c d\varepsilon_{1a}/d\kappa$
is defined by $J(\kappa_0)$-integral. Fig. 4 illustrates calculated $J$-value
as a function of $\kappa_0$. We see long enough plateau beginning from $\kappa_0$
slightly  more than $\kappa_c$. It is convenient to select such $\kappa_0$ 
that $\varepsilon_{1a}(\kappa_1)=0$, where we denote this special value of $\kappa_0$- set
as $\kappa_1$. It is easy to see that 
\begin{equation}
\kappa_1 = \kappa_0 + {\varepsilon_{1a}(\kappa_0)\over {d\varepsilon_{1a}\over d\kappa} (\kappa_0)}  .
\end{equation}   
where $\kappa_0$ is any value of $\kappa$ in the range of linearity of $\varepsilon_{1a}$ function.
According to our calculation for the silicon (110) plane $\kappa_1 \approx 0.13$.
Now we can wrote the following relation:
\begin{equation}
 \bar{ \varepsilon}_{1a}={AU_0^{1\over 4}J_p \over 2^{7\over 4}\sqrt{\pi}E_0^{1\over 4}{\cal{E}}_{max} d^{1\over 2} X_0^{1\over 2}} 
({\kappa \over \kappa_c} -{\kappa_1 \over \kappa_c})
\end{equation}
Here we denote the linear approximation of $\varepsilon_{1a}$ (see Eq.(9))  as $\bar {\varepsilon}_{1a}$
and the plateau value of $J$ as $J_p$ . 

We calculated also value $\varepsilon_{1b}$ (see Eq.(25) ) for different crystals.
 According to these calculations  relation $\varepsilon_{1b}/\varepsilon_{1a}$
is approximately equal to 0.41- 0.37 at $R/R_c= 1.5-30 $, correspondingly. We think that we can decide
this relation approximately constant in this region of radii and equal to 0.39.
Then we can write the final equation for the probability $\varepsilon_1$:
\begin{equation}
 \bar{ \varepsilon}_{1} \approx  {1.39AU_0^{1\over 4}J_p \over 2^{7\over 4}\sqrt{\pi}E_0^{1\over 4}{\cal{E}}_{max} d^{1\over 2} X_0^{1\over 2}} 
({\kappa \over \kappa_c} -{\kappa_1 \over \kappa_c})
\end{equation}
 Analogously, here $\bar{ \varepsilon}_{1}$ denotes the linear approximation of $\varepsilon_{1}$. 
 
Fig.5 illustrates behavior of the relations $\varepsilon_1/ \bar{\varepsilon}_1$ and
$\varepsilon_{1a}/\bar{\varepsilon}_{1a}$ as functions of variable $X=(\kappa-\kappa_1)/\kappa_c$.
One can see that in the region from $X=0.2$ till $\approx 2.5$ these relations are constant
with a good enough accuracy. Thus, Eq.(34) allows one to describe by universal way
the probability of volume capture in the area of interest to us.  
Table 1 contains the parameters of some single crystals for calculation of $\varepsilon_1$-value. 

Fig. 6 illustrates the differential over energy distributions of captured protons
which were calculated with the help of Eqs. (23) and (25).
We see that these distributions become more wide with the increasing of
the bending radius.

\subsection{Channeling of captured particles}
The channeling of captured particles we illustrate for proton energy 400 GeV
and 10 m bending radius for the (110) plane of a silicon single crystal.
The similar simulations based on Monte Carlo calculations for these parameters
presented in Ref.{\cite{ST}}. and hence, we can compare results  which were
carried out by different methods. 
Eq.(26) allow one to calculate the flux of particles which were captured in the channeling
regime. Fig.7  shows  behavior of the flux as a function of the thickness of a
crystal.  
The curves 1 and 2 (see fig.7b) present the speed of dechanneling as a function
of variable bending angle for cases without and with consideration of multiple scattering
in a silicon crystal. In this figure  the curve 3 presents the distribution function
of dechanneling particles on the exit of a single crystal.
The total number of captured particles is $\approx 0.057$.
The same number is $\approx 0.064$ accordingly to Ref.{\cite{ST}}.

\subsection{Total scattering curve}
Thus, the total scattering curve (with taking account the volume capture process) 
represents the sum of the third curves: 

i) Pure volume reflection curve;

ii) the curve of the dechanneling fraction;

iii) the curve of the channeling fraction.

The resulting curve (for 400 GeV protons and 10 m bending radius of (110) plane
silicon single crystal) is shown in Figs. 8 and 9   in linear and logarithmic scales,
correspondingly.
The curve of channeling fraction was calculated with accordance of Eq.(31).
The semiperiod in Eq.(31) was chosen at the condition that in the moment of  time $t_1$
the transversal velocity is maximal and in the moment of time $t_2=t_1+\tau$
this velocity is minimal.    

One can see that in accordance with the distribution function most particles of channeling 
fraction are groped near critical channeling angle. This fact is due to
firstly, the large number of particles have high transversal energy and secondly,
the time of stay in state with small transversal velocity is considerable larger
than with the time of state with the large velocity.
It should be noted that experimental observation of two peaks of the channeling curve
is practically impossible because of multiple scattering in detectors.

The important characteristics of volume reflection of a beam is the efficiency
of the process which defined as a partition of beam which contained 
in $\pm 3\sigma_{vr}$ around of mean angle of reflection $\alpha_{m}$, where $\sigma_{vr}$ is the mean square of angle distribution
of a scattered beam. The calculated for conditions as in Figs. 8 and 9 the efficiency
is close to 0.98.  This is in agreement with the experimental data \cite{WS2}.
From this result follows that at above mentioned definition of the efficiency $\varepsilon$
the total inefficiency ( $1- \varepsilon$) should be significantly less than the probability
of volume capture $\varepsilon_1$ due to a valuable  fraction of particles, which
quickly return from channeling  in  underbarier motion (see Fig. 7).    

\section{Discussion}
As mentioned above our consideration is valid for a small enough bending radii
of  single crystals. It is clear that for large bending radii the energy gap
$\delta E$ becomes very small. Because of this, the particle energy losses can be
significantly exceed value $\delta E$, and, consequently, the possibility
of volume capture will realized for long enough part of trajectory of  particles, 
instead of short distance from $x_1$ till $x_3$ (see Fig. 1).   

Let us estimate the area of application of our analytical
description. Taking into account that distribution of energy losses is described by normal
one we can write the following condition for validity of obtained in the paper equations:
\begin{equation}      
   \sigma_{E,max} < \delta E/4
\end{equation} 
where $\sigma_{E,max}$ is the maximal value of $\sigma_E$ at fixed $\kappa$-parameter.
For the (110) silicon plane and proton energy equal to 400 GeV $\delta E/4 [eV] \approx 20/R[m]$
and $\sigma_{E,max} \sim 1$ eV and we got that our consideration is valid till 
$R \sim 20$ m (or for $R/R_c \sim 30$). The area of validity is extended with the
increasing of particle energy.
It is important to note that the range $R/R_c\approx 10 -20$ is more preferable
for utilization of the volume reflection process  on accelerators. Really,
at these values of radii the mean angle $\alpha_m$ of the process is close to
maximal one \cite{MV}  and the volume capture process is suppressed in comparing with
more large bending radii.

In this paper for small enough bending radii we got simple relation for probability
of volume capture $\varepsilon_1$ in different single crystals. According to this relation
$\varepsilon_1$ is proportional to $E_0^{-{1\over4}}(R/R_c-R_1/R_c)$. It means
that $\varepsilon_1$ is a weakly dropping function of particle energy 
at fixed $R/R_c$. For testing of this dependence we carried out Monte Carlo
calculations of $\varepsilon_1$ at different radii and energies. 
The results of these calculations are described by the function which
is proportional to $E_0^{-0.2}(R/R_c-0.7)$. These dependence is close to obtained
one in this paper. We plan to continue Monte Carlo calculations with the aim of
obtaining large statistics and  hope to find the degree at $E_0$ which gives
the best description of volume capture.

Process of volume capture was investigated also in Ref.{\cite{CK}}. Here 
the probability of volume capture is $\varepsilon_1=R\theta_{ch}/L_d$, where
$L_d$ is the dechanneling length, which is proportional to particle energy. It means that $\varepsilon_1$ is proportional
to $E_0^{-{1\over 2}} R/R_c$.The paper \cite{ST} contains the criticism of this
relation. Authors \cite{ST} give the conclusion about incorrectness of the relation and
 try to correct its by introducing of  a nuclear dechanneling length. 
This correction allow one to get values of probability
 close to Monte Carlo calculations (for energy 400 GeV)  but conserve the previous energy dependence.
\section{Conclusion}
The main results of our investigation of volume reflection of relativistic particles
are:

1) the simple analytical relations for probability of volume capture
are obtained;

2) propagation of volume captured fraction inside a single crystal is
considered;

3) the analytical method for calculation of the angle scattered distribution of particles, taking into account different
processes  are developed;

4) presented here  equations allow one to find efficiency of the volume reflection process
but  analysys of this problem is beyond of the scope of this paper.
  
\section{Asknowledgments}
Authors would like to thank all the participants of  CERN experiment H8RD22  for collaboration.
 
This work was partially supported by Grant No. INTAS-CERN 05-103-7525 and
Russian Foundation for Basic Research Grant No. 08-02-01-453.

\newpage
\begin{center}
\begin{table}
\begin{tabular}{|c|c|c|c|c|c|c|c|c|}
\hline
$Crystal$ & $Z$ & $X_0,cm$ & $d,A$ & $U_0,eV$ & ${\cal{E}}_{max},eV/A$ & $\kappa_c$ & $\kappa_1$ & $J_p$\\   
\hline
$C(diamond)$ & $6$ & $12.14$ & $1.26$ & $23.52$ & $79.70$ & $0.234$ & $0.198$ & $1.71$\\ 
$Si$ & $14$ & $9.38$ & $1.92$ & $21.38$ & $60.0$ & $0.186$ & $0.13$ & $1.49$\\
$Ge$ & $32$ & $2.28$ & $2.00$ & $40.67$ & $111.46$ & $0.183$ & $0.123$ & $1.44$\\
$W$ & $74$ & $0.383$ & $2.23$ & $138.34$ & $490.3$ & $0.126$ & $0.0085$ & $1.29$\\
\hline
\end{tabular}
\caption{Parameters of single crystals for calculation of volume capture probability (see Eq.(34)).
Potentials for diamond and silicon are taken from x-ray measurements\cite{CW,Ti} and 
for germanium and tungsten Molier potential was used.}
\end{table}
\end{center}

\begin{figure} 
\begin{center}
\parbox[c]{14.5cm}{\epsfig{file=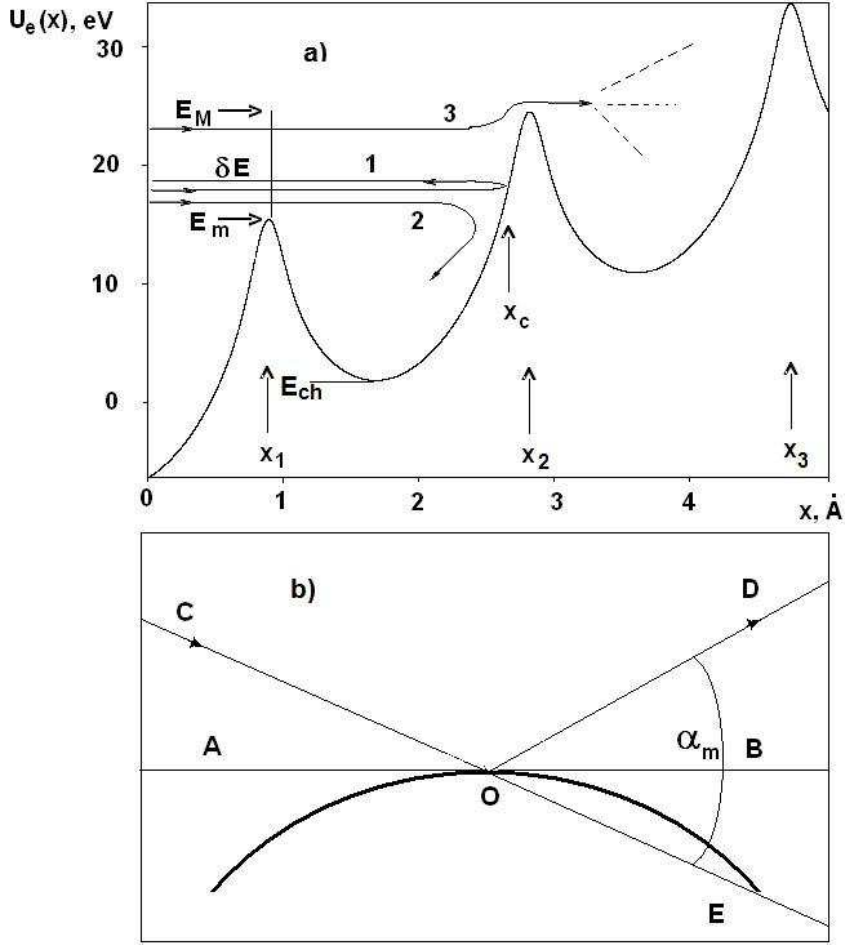,width=12cm}}
\parbox[c]{15cm}{\caption{
Scheme of volume reflection and volume capture processes: a) the effective potential  of
bent crystallographic planes, $x$ is the transversal coordinate.
$E_M$ and $E_m$ are the transversal energies corresponding to 
neighbouring local maxima of the potential; $E_{ch}$ is one of transversal energies
corresponding to local minimum of potential; $x_1, x_2, x_3$ are the transversal
coordinates of the local maxima; the curves 1,2,3  reflect the different possibilities
for moving particles.  $x_c$ is the critical coordinate for volume reflected particles;
b) geometric relations of the process.
For additional information, see the text.          
    }}
\end{center}
\end{figure}

\begin{figure} 
\begin{center}
\parbox[c]{14.5cm}{\epsfig{file=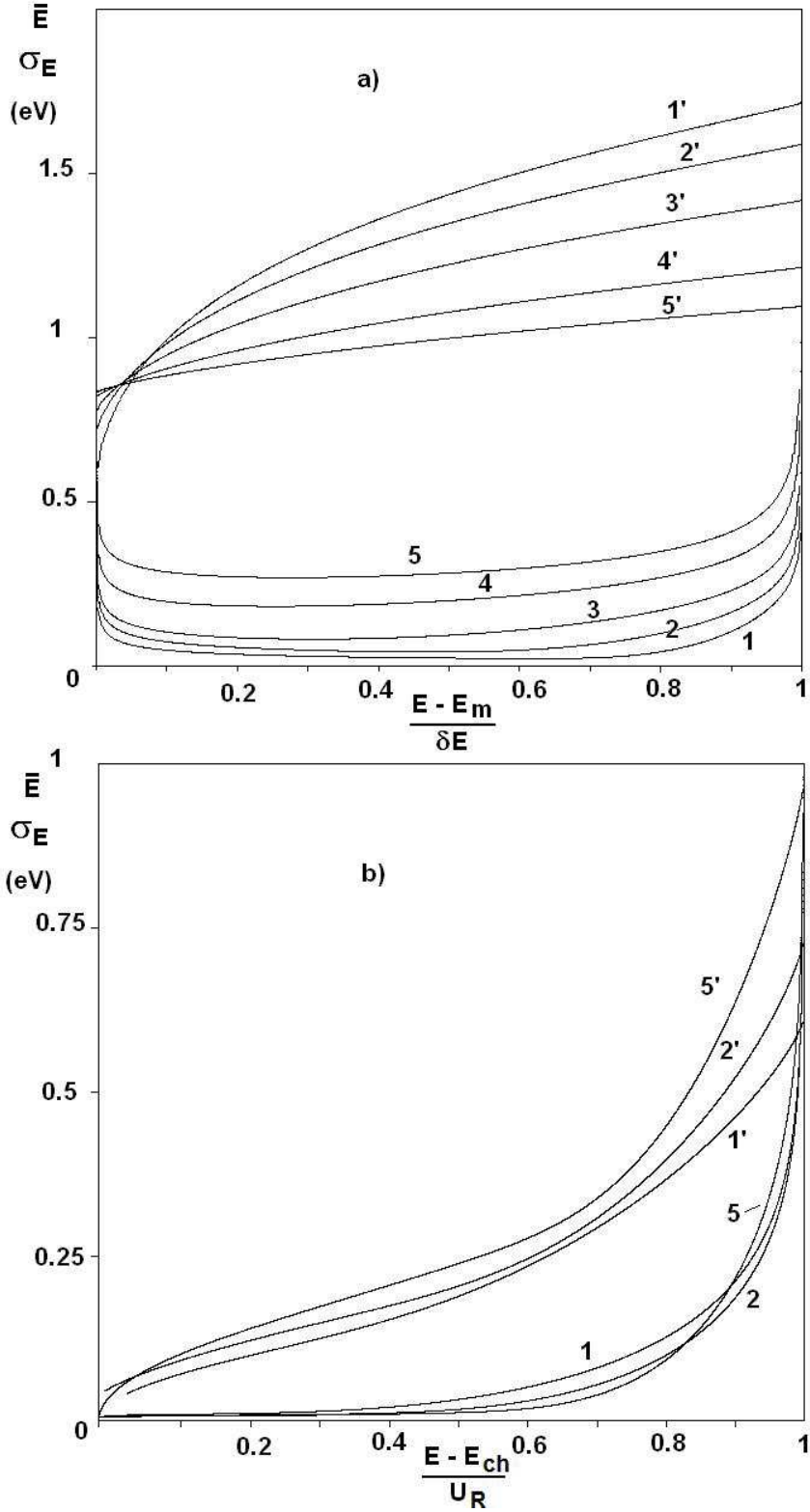,width=12cm}}
\parbox[c]{15cm}{\caption{
   Losses of transversal energies  $\bar{E}$ and $\sigma_e$ (in the (110) silicon plane) for volume reflection (a)
and for channeling regime (b) as functions of variables $(E-E_m)/\delta E$ 
and $(E-E_{ch})/U_R$, correspondingly. The numbers $1-5$ near curves are equal to value of 
$\kappa$-parameter and correspond to $\bar{E}$-quantities. Analogously the
number $1'-5'$ are equal to value of $\kappa$-parameter and correspond to 
$\sigma_E$-quantities. The values $\bar{E}$ in (b) are enlarged in three times for clearness.
The particle energy is equal to 400 GeV. 
     }}
\end{center}
\end{figure}

\begin{figure} 
\begin{center}
\parbox[c]{14.5cm}{\epsfig{file=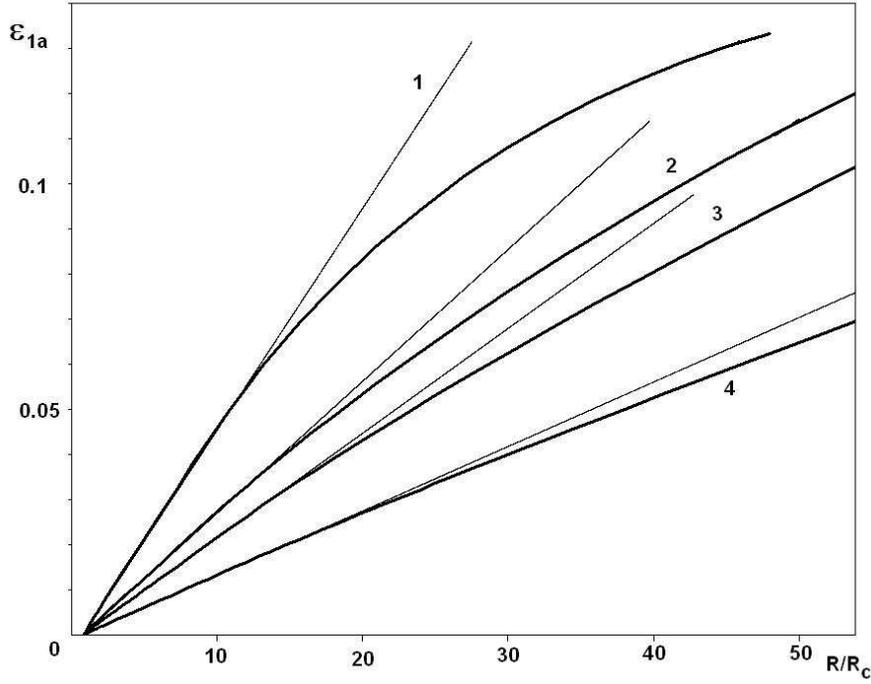,width=12cm}}
\parbox[c]{15cm}{\caption{
$\varepsilon_{1a}$-value for the (110) silicon plane as a function of relation $R/R_c$ for different energies
 50 GeV (1), 400 GeV (2), 1000 GeV (3), 7000 GeV (4). The straight lines
near curves are linear approximations (see Eq. (19)). 
    }}
\end{center}
\end{figure}

\begin{figure} 
\begin{center}
\parbox[c]{14.5cm}{\epsfig{file=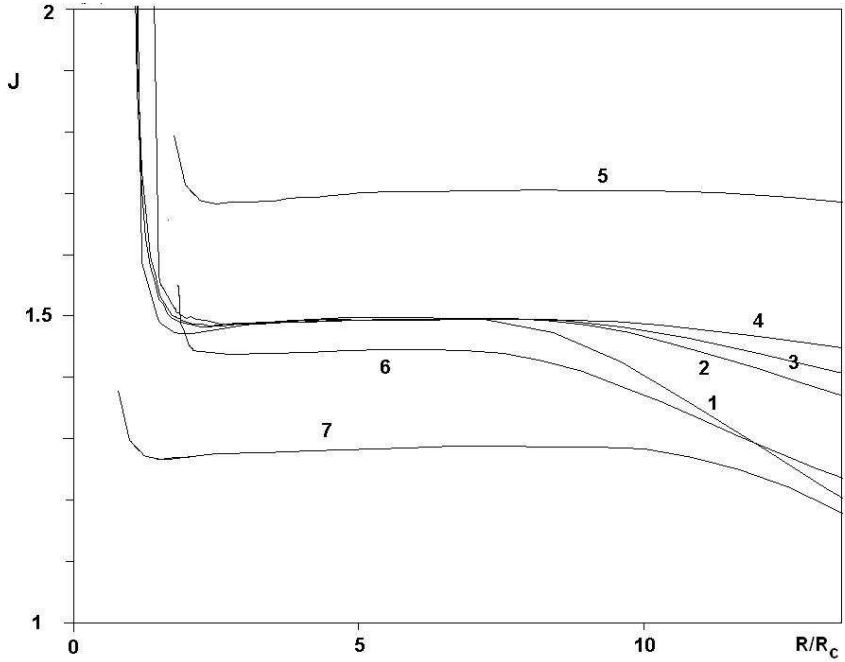,width=12cm}}
\parbox[c]{15cm}{\caption{
J-value as a function of relation $R/R_c$ for different conditions.
Curves 1,2,3,4 are for silicon single crystal and proton energies 50, 400, 1000 and
7000 GeV, respectively. Curves 5,6,7 are for diamond, germanium and tungsten 
single crystals and proton energy equal to 400 GeV.
    }}
\end{center}
\end{figure}

\begin{figure} 
\begin{center}
\parbox[c]{14.5cm}{\epsfig{file=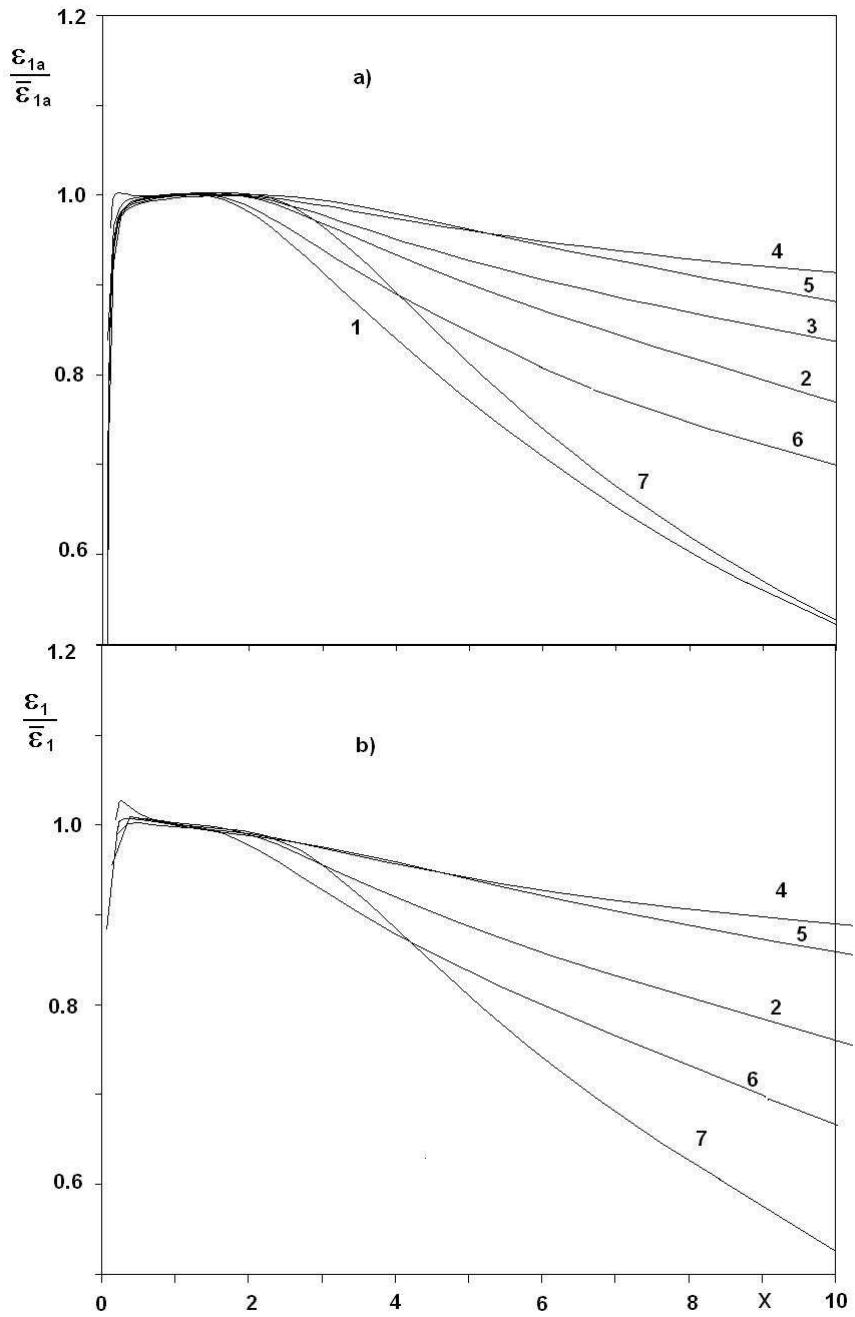,width=12cm}}
\parbox[c]{15cm}{\caption{
The relations $\varepsilon_{1a}/\bar{\varepsilon_{1a}}$ (a) and
$\varepsilon_1/\bar{\varepsilon_1}$ (b) as functions of X-value (see text).
Number near curves are the same as in Fig. 4. 
   }}
\end{center}
\end{figure}

 \begin{figure} 
\begin{center}
\parbox[c]{14.5cm}{\epsfig{file=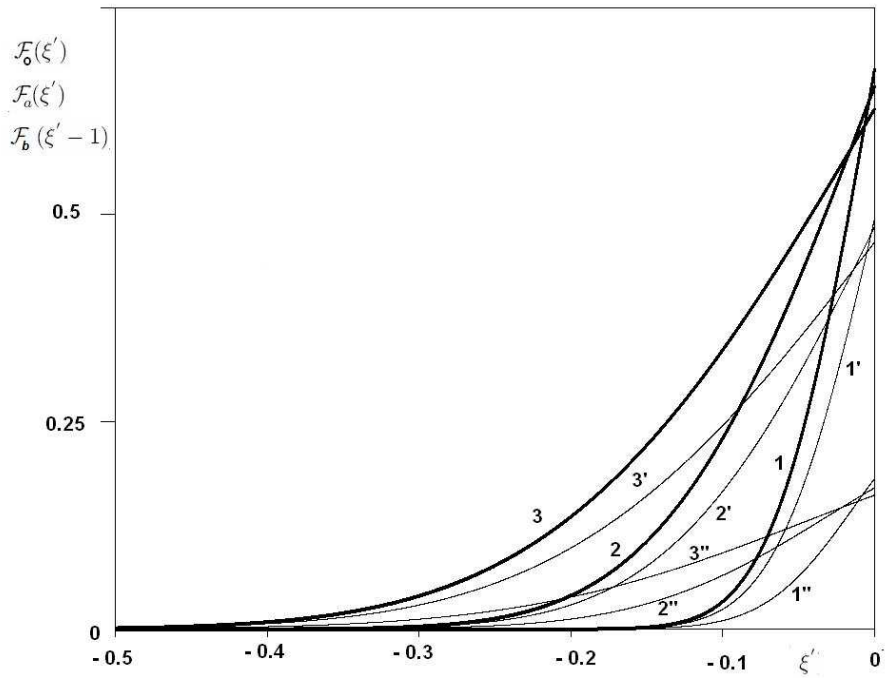,width=12cm}}
\parbox[c]{15cm}{\caption{
The distributions ${\cal{F}}_0, {\cal{F}}_a, {\cal{F}}_b$ as functions of
$\xi$-parameter for bending radii equal to 5, 10, 15 m (curves 1,2,3, respectively).
${\cal{F}}_0={\cal{F}}_a+{\cal{F}}_b$. The curves labeled as $'$ and $''$ correspond to
${\cal{F}}_a, {\cal{F}}_b$ -distributions and the thick curves to ${\cal{F}}_0$ one.
   }}
\end{center}
\end{figure}
 \begin{figure} 
\begin{center}
\parbox[c]{14.5cm}{\epsfig{file=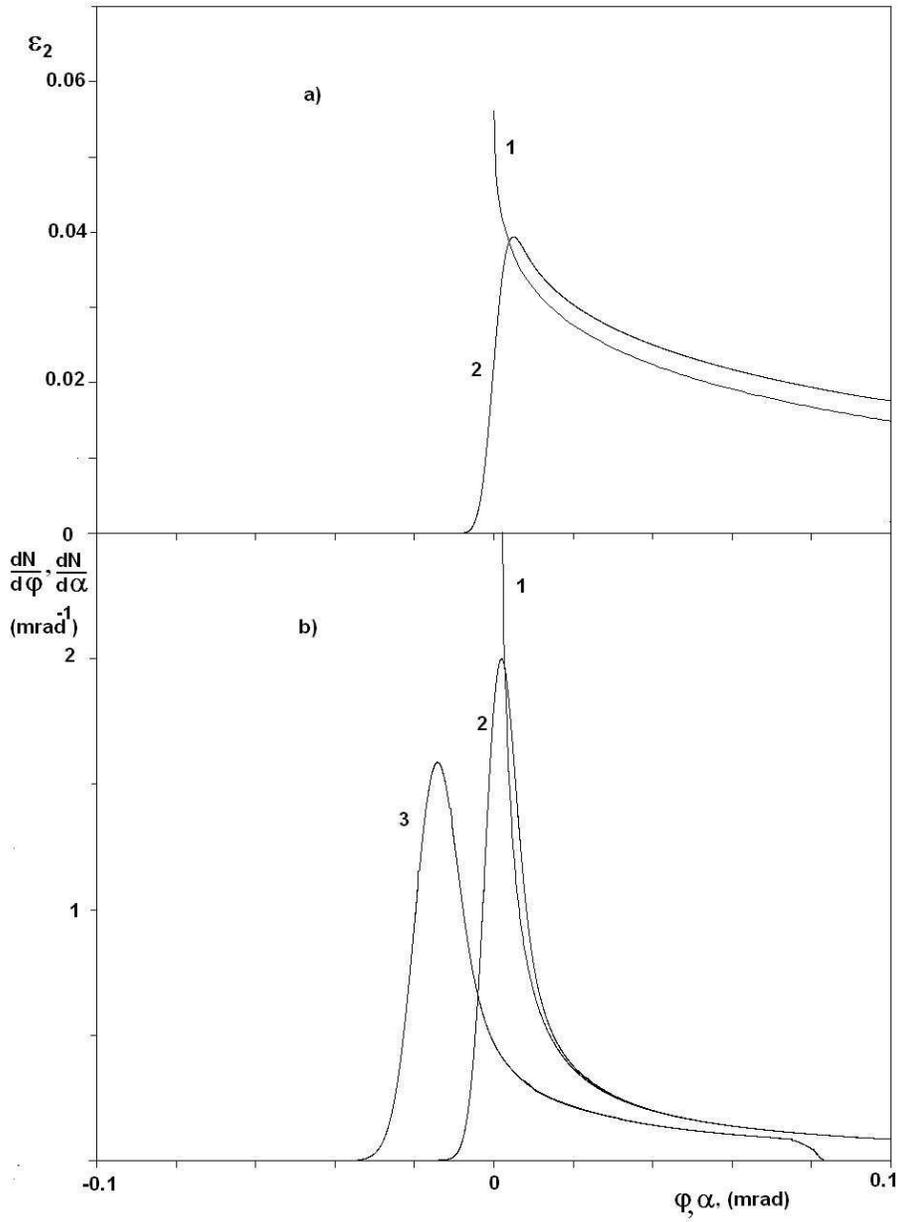,width=12cm}}
\parbox[c]{15cm}{\caption{
    Intensity of channeling fraction (a) and differential particles losses due to dechanneling (b)
 as  functions of angle $\varphi$. Curves 1 and 2 correspond to considerations without and with
multiple scattering. Curve 3 corresponds to angle distribution (over $\alpha$-angle)
dechanneling fraction
on the entrance of  a single crystal.   
   }}
\end{center}
\end{figure}
 \begin{figure} 
\begin{center}
\parbox[c]{14.5cm}{\epsfig{file=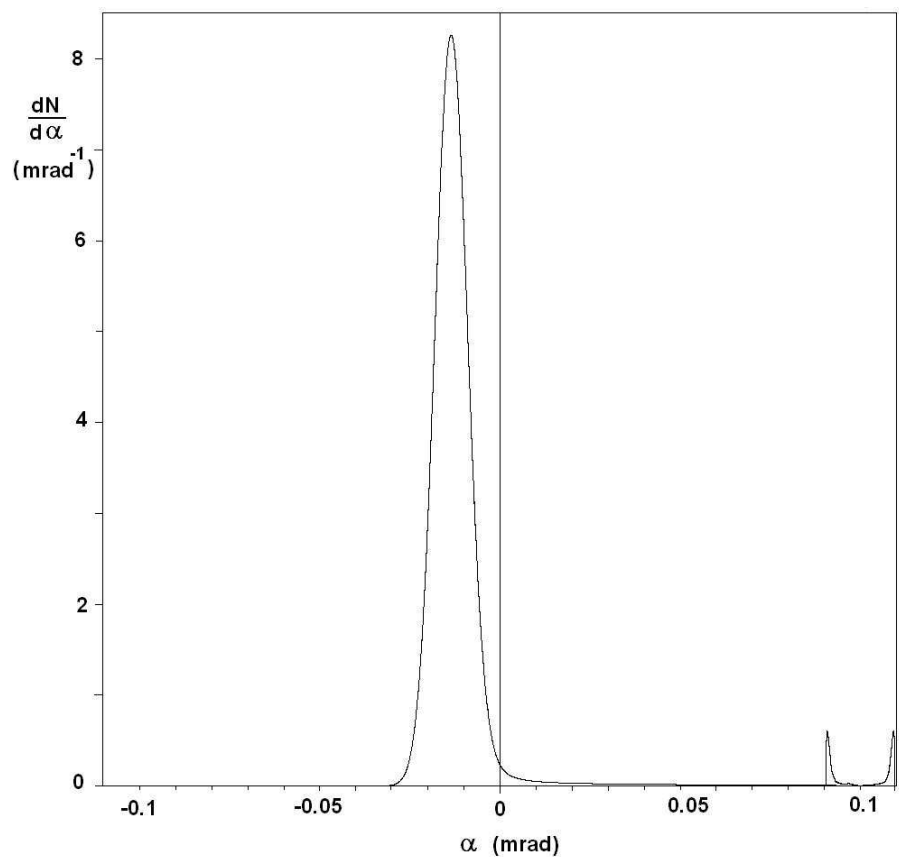,width=12cm}}
\parbox[c]{15cm}{\caption{
The resulting distribution of scattered particles at volume reflection taking into account influence 
of volume capture. The energy of protons is equal to 400 GeV, bending radius of 2-mm silicon
single crystal is equal to 10 m. 
   }}
\end{center}
\end{figure}
 \begin{figure} 
\begin{center}
\parbox[c]{14.5cm}{\epsfig{file=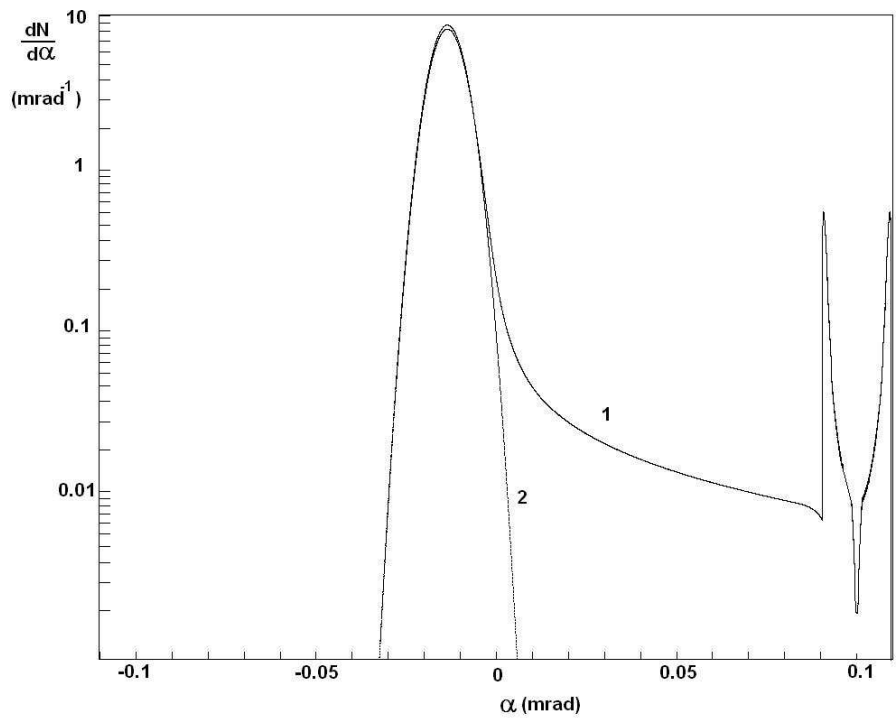,width=12cm}}
\parbox[c]{15cm}{\caption{
The curve 1 as the same as in Fig. 8 but y-scale is logarithmic.
The curve 2 is calculated without taking into account the volume capture.  
 }}
\end{center}
\end{figure}

\end{document}